\documentstyle[12pt]{article}
\def\m{\mu}\def\n{\nu}

\def\e{\epsilon}\def\be{\begin{equation}}
\def\ee{\end{equation}}\def\p{\partial}\def\ber{\begin{eqnarray}}
\def\eer{\end{eqnarray}}
\begin{document}\begin{center}
{\Large\bf A Note on Duality Symmetry in Nonlinear Gauge Theories }
\vskip 1 true cm{\bf Rabin Banerjee}
{\footnote{ On leave from S.N.Bose National Centre for Basic Sciences, Calcutta, 
India; e-mail: rabin@newton.skku.ac.kr;  rabin@bose.res.in}}
\vskip .8 true cm
BK21 Physics Research Division and Institute of Basic Science,

SungKyunKwan University,

Suwon 440-746, Republic of Korea.
\end{center}
\vskip 1 true cm

{\centerline{\bf{Abstract}}}
\bigskip

An intriguing connection, based on duality symmetry,
 between ordinary (commutative) Born-Infeld type theory and
non-commutative Maxwell type theory, is pointed out. Both discrete 
as well as continuous duality transformations are considered and their
implications for self duality condition and Legendre transformations are 
analysed.
 
\newpage

The study of duality symmetry in different contexts has led to 
new results and important insights {\cite{book}}. In the realm of field theory,
perhaps the most widely studied example (apart from the simple Maxwell theory)
is the Born-Infeld theory or variants of it, which go by the common name of 
nonlinear electromagnetism {\cite{review}}.
Recently such investigations have been extended to non-commutative field 
theories, i.e.; field theories defined on non-commutative spaces. Examples are
the non-commutative Maxwell and non-commutative Born-Infeld theories \cite{
GRS, SJR,
SR, A, ABT}.
Duality rotations in these theories are discussed by means
of  Legendre transformations \cite{GRS, SJR, SR} or by considering the decoupling 
limit of D-brane effective actions in the slowly varying field approximation 
\cite{A}. In \cite{ABT}, use has been made of the duality symmetry to discuss
modified dispersion relations in non-commutative electrodynamics.

In this paper we discuss both discrete and continuous 
duality symmetry transformations. 
We consider generalised versions of ordinary 
Born-Infeld type lagrangians and non-commutative
Maxwell type lagrangians. Both these theories are shown to be duality invariant under
discrete transformations. In extending the discrete symmetry to the continuous one, the
starting point is the self duality criterion
given in \cite{GZ} and reviewed in \cite{GZ1}.
The solution of this criterion  leads to a remarkable 
similarity among these generalised theories. It is found, in both cases, that
the ratio of the coefficients of the nonlinear terms must be four. Under this
condition the generalised theories reduce to the ordinary Born-Infeld lagrangian
and the commutative equivalent of the standard
non-commutative Maxwell theory obtained by an application
of the Seiberg-Witten map \cite{GRS}. We also show that, although self duality
condition is satisfied only for a specific value (i.e. 4) of the ratio. the invariance
under Legendre transformations, which is another way of looking at duality,
remains valid for any ratio, in both the models.
The implications of including quantum effects are briefly discussed.

The usual Born-Infeld lagrangian is expressed in terms of the field tensor
$F_{\mu\nu}$ as,
\be
L= -\frac{1}{g^2}\Big(\sqrt{- det(\eta_{\mu\nu}+g F_{\mu\nu}})-1\Big)
\label{BI0}
\ee
which, in the leading order, simplifies to,
\be
L= -\frac{1}{4} F^2 - \frac{g^2}{32}(F^2)^2 +\frac{g^2}{8}(FFFF){^\m\,\, _\m}
\label{BI}
\ee
where $F^2=(F_{\mu\nu} F^{\mu\nu})$ and the matrix notation,
\be
(AB){^\m\,\,_\m}=A^{\m\n}B_{\n\m}
\label{notation}
\ee
will be consistently used.

Defining,
\be
^* G = 2\frac{\p L}{\p F}\,\,\,;\,\,\, ^* G_{\mu\nu}=\frac{1}{2}
\e_{\mu\nu\lambda\rho}G^{\lambda\rho}
\label{dual}
\ee
the equations of motion and the Bianchi identities get exprssed as,
\be
\p_\mu\, ^*G^{\mu\nu} = 0\,\,\,;\,\,\, \p_\mu\, ^*F^{\mu\nu} = 0 
\label{eqmotion}
\ee
Then it is known that this set of equations is preserved under the 
discrete electric-magnetic
duality transformation,
\be
F\rightarrow G\,\,\,;\,\,\, G\rightarrow -F
\label{duality}
\ee
which may also be extended to a continuous $SO(2)$ rotation,
\be
\left(\begin{array}{c}
G'\\
F'
\end{array}\right)
=
\left(\begin{array}{clcr}
\cos\lambda & - \sin\lambda \\
\sin\lambda & \cos\lambda
\end{array}\right)
\left(\begin{array}{c}
G\\
F
\end{array}\right)
\label{matrix}
\ee
whose infinitesimal versions are given by,
\be
\delta G = -\lambda F\,\,\,;\,\,\, \delta F = \lambda G
\label{inf}
\ee

Note that the discrete transformation corresponds to taking $\lambda=\frac{\pi}{2}$.
Since $G$ is a function of $F$, an essential ingredient for the duality symmetry to
be self consistent is to preserve the stability of the definition (\ref{dual}). This
is ensured, for the continuous
duality rotation, by the consistency condition \cite{GZ, GZ1},
\be
G ^* G + F ^* F = 0
\label{consistency}
\ee

It is obvious that invariance under the continuous symmetry transformation 
would imply invariance under the discrete transformation. The converse, however,
need not be true; i.e. it may be possible
to have a theory that has discrete duality invariance, but lacks the
continuous symmetry. In other words, the consistency condition (\ref{consistency})
is sufficient, but not necessary, for requiring the theory to be invariant under 
discrete duality transformation. To elaborate on this point, we consider a general
Born-Infeld type lagrangian with arbitrary coefficients for the nonlinear terms.
Moreover, since the coupling can be scaled, only the ratio of the coefficients is
significant. We thus consider the following lagrangian,
\be
L= -\frac{1}{4} F^2 - g^2(F^2)^2 +a g^2 (FFFF){^\m\,\, _\m}
\label{BItype}
\ee
When the ratio $a=4$, it reduces to the usual Born-Infeld theory. Using (\ref{dual}), 
we find,
\be
^* G_{\mu\nu}=-F_{\mu\nu} - 8g^2 F^2 F_{\mu\nu} - 8 a g^2 (FFF)_{\mu\nu}
\label{dual1}
\ee
from which one also obtains,
\be
G=-^{**} G =\, ^*F + 8g^2 F^2 ({^*}F) + 8 a g^2\, {^*}(FFF)
\label{dual2}
\ee
Obviously the equations of motion together with the Bianchi identities will be
preserved under the discrete duality map (\ref{duality}). 
In order to be self consistent, it is however essential to see the 
stability of the relations (\ref{dual1})
and (\ref{dual2})
 under this mapping.
Consider therefore the  effect of $F\rightarrow G$ on (\ref{dual1}). Then it follows,
\be
^*G\rightarrow -G - 8g^2 G^2 G - 8 a g^2 (GGG)
\label{gvariation}
\ee
Putting the value of $G$ from (\ref{dual2}) in (\ref{gvariation}), we get,
\be
^*G\rightarrow -\,^*F - 8ag^2 [^*(FFF) + ({^*F}{^*F}{^*F})]
\label{gvariation1}
\ee 
Using the identity,
\be
^*(FFF) + ({^*F}{^*F}{^*F}) = 0
\label{identity}
\ee
the above relation simplifies to $^*G\rightarrow -^*F$, thereby reproducing the
second transformation in (\ref{duality}) and demonstrating the consistency.
Thus the lagrangian with an arbitrary parameter manifests the symmetry under
the discrete duality map. 

For the continuous case, recourse has to be taken to the 
condition (\ref{consistency}). Now a straightforward algebra, using (\ref{dual1})
and (\ref{dual2}), yields, 
\be
G^*G = -F^*F - 16 g^2 [ (F^2)F^*F + a ^*F (FFF)]
\label{11}
\ee
which, exploiting the identity,
\be
F^2 (F^*F) = -4 ^*F(FFF)
\label{identity1}
\ee
simplifies to,
\be
G^*G + F^*F = -16 g^2 (a - 4) ^*F(FFF)
\label{22}
\ee
The consistency condition (\ref{consistency}) is satisfied provided $a=4$ in which
case the original Born-Infeld lagrangian is obtained.

The inference is that the general lagrangian with an arbitrary parameter is 
duality invariant under discrete transformations, but only the Born-Infeld manifests
the continuous symmetry{\footnote{Of course there are other variants of  
nonlinear lagrangians (which may \cite{HKS} or may not \cite{GR}
have the Maxwell weak field limit) that are self dual,
 but here our interest concerns the
specific type (\ref{BItype}) with only quartic $F$-terms, which reduce to the
Maxwell theory for weak fields}. 

These considerations are
now extended to nonlinear electromagnetism defined on noncommutatice spaces. In
reality, the commutative equivalents of such theories will be analysed.
This is obtained from (\ref{BItype}) by replacing, in the nonlinear sector, one
of the field strengths by a constant 2-index object $\theta^{\mu\nu}$, which 
characterises the noncommutativity parameter. Thus we get the form,
\be
L= -\frac{1}{4} F^2 -  (\theta F)(F^2) + b  (\theta FFF){^\m\,\, _\m}
\label{NCEMtype}
\ee
where the coupling has been absorbed and the ratio is given by the parameter
$b$ to indicate that, in general, it can be different from that appearing in 
(\ref{BItype}). Equation (\ref{NCEMtype}) defines a commutative equivalent of 
Maxwell type theory defined in non-commutative space, upto the first order in the
non-commutative parameter. 
It is the most general lagrangian constructed out of $F^{\mu\nu}$ and $\theta^{\mu\nu}$,
up to $O(\theta)$, which,
in the limit of vanishing $\theta$, reduces to the usual Maxwell
theory. At this point the ratio $b$ is completely arbitrary.

As was done for (\ref{BItype}), we study the consequences of duality symmetry on
(\ref{NCEMtype}). First, the discrete transformations are discussed. From the 
definition (\ref{dual}), it follows that,
\be
^*G = -F - 4(\theta .F) F - 2\theta F^2 - 2b (\theta FF +F\theta F + FF\theta)
\label{ncdual}
\ee
and,
\be
G =\, ^*F + 4(\theta .F)\, ^*F + 2\,^*\theta F^2 + 2b\,
 ^*(\theta FF +F\theta F + FF\theta)
\label{ncdual1}
\ee
Now it is clear that enforcing $F\rightarrow G$ in (\ref{ncdual}) does 
not lead to $G\rightarrow -F$ due to the presence of the $\theta$-term.
An appropriate transformation for $\theta$ has to be defined, which
is given by,
\be
\theta\rightarrow \,^*\theta
\label{theta}
\ee
 It is suggested
by the fact that since (\ref{NCEMtype}) was obtained from (\ref{BItype})
by a formal replacement of a $F$ by $\theta$, their transformation properties
should be similar. In the lowest order the map $F\rightarrow G$ reduces, on
using (\ref{ncdual1}), to $F\rightarrow ^*F$, leading to the above transformation for
$\theta$. With the combined transformations $(F\rightarrow G, \theta
\rightarrow \,^*\theta)$, the change in $^*G$ becomes,
\be
^*G\rightarrow -G -4(^*\theta .G)G - 2^*\theta G^2 - 2b (^*\theta GG
+G^*\theta G + GG ^*\theta)
\label{33}
\ee
Substituting the value of $G$ from (\ref{ncdual1}), we get,
\be
^*G\rightarrow - ^*F -2b ^*(\theta FF + F\theta F + FF\theta) 
-2b (^*\theta {^*F} {^*F} + ^*F {^*\theta}{^* F} + {^*F}{^*F}{^*\theta})
\label{44}
\ee
Using identities similar to (\ref{identity}) (which is also valid when one of
the $F's$ is replaced by $\theta$), there is a pairwise cancellation of all the
$\theta$-terms, yielding the cherished transformation law (\ref{duality}). Thus the 
lagrangian (\ref{NCEMtype}) manifests the discrete duality symmetry, once the 
transformation on $\theta$ given by (\ref{theta}) is taken.

Now the continuous duality rotations will be considered. In the presence of 
additional variables (in this case, $\theta$), the self dual condition 
(\ref{consistency}) is modified as \cite{GZ1, A},
\be
\frac{\lambda}{4}(G^*G + F^*F) = -\frac{\p L}{\p\theta}\delta\theta
\label{variation}
\ee
where the infinitesimal change in $\theta$ follows from (\ref{theta}),
\be
\delta\theta = \lambda ^*\theta
\label{inftheta}
\ee
In fact the complete rotation symmetry of $\theta$ may be expressed exactly 
by the same matrix appearing in (\ref{matrix})
\be
\left(\begin{array}{c}
^*\theta'\\
\theta'
\end{array}\right)
=
\left(\begin{array}{clcr}
\cos\lambda & - \sin\lambda \\
\sin\lambda & \cos\lambda
\end{array}\right)
\left(\begin{array}{c}
^*\theta\\
\theta
\end{array}\right)
\label{matrix1}
\ee
Expectedly, as before, the discrete transformation (\ref{theta}) is obtained for
$\lambda=\frac{\pi}{2}$.

From (\ref{ncdual}) and (\ref{ncdual1}) we obtain,
\be
G ^*G = -F ^*F - 4\Big(F^2(\theta ^*F) + (2-\frac{3b}{4})(F ^*F)(\theta F)\Big)
\label{55}
\ee
Also, from the lagrangian, it follows,
\be
\frac{\partial L}{\partial\theta^{\alpha\beta}}= - \Big(F^2 F_{\alpha\beta}
+b (FFF)_{\alpha\beta}\Big)
\label{66}
\ee
Putting in all these expressions in the condition (\ref{variation}), yields,
\be
F^2(\theta ^*F) + (2-\frac{3b}{4})(F ^*F)(\theta F)= -\Big(F^2 F_{\alpha\beta}
+b (FFF)_{\alpha\beta}\Big) ^*\theta^{\alpha\beta}
\label{77}
\ee
Finally, using the identity,
\be
\Big( ^*\theta FFF\Big){^\alpha\,\,_\alpha}=\frac{1}{2}F^2(\theta ^*F)
- \frac{1}{4}(F ^*F)(\theta F)
\label{identity0}
\ee
equation (\ref{77}) simplifies to,
\be
(b-4)\Big(F^2(\theta ^*F)
+ (F ^*F)(\theta F)\Big) = 0
\label{constant}
\ee
so that $b=4$.

Remarkably, the same ratio, as in the case of the Born-Infeld example, is obtained.
For the sake of comparison we recall that the non-commutative version of the 
Maxwell lagrangian, expressed in terms of its commutative equivalent by using the 
Seiberg-Witten map, is given by \cite{GRS},
\be
L= -\frac{1}{4} F^2 +  \frac{1}{8}(\theta F)(F^2) - \frac{1}{2}
  (\theta FFF){^\m\,\, _\m}
\label{NCEM}
\ee
where terms up to the leading order in $\theta$ have only been retained.
Since the ratio of the coefficients of the correction terms is 4(including
the correct sign), this lagrangian is equivalent to (\ref{NCEMtype}) with
$b=4$.

We conclude that the general non-commutative theory has only discrete duality 
symmetry. For the continuous symmetry to hold, the ratio is determined uniquely and
agrees with the  non-commutative Maxwell
theory obtained by an application of the Seiberg-Witten
map.
In fact it is possible to use this analysis to invert the usual argument of exploiting
the Seiberg-Witten map to obtain the commutative equivalent of non-commutative
Maxwell theory. One starts from a general lagrangian like (\ref{NCEMtype}) and demands
invariance under duality rotations. This fixes the ratio. Now the map can be
derived that connects this theory with the Maxwell theory defined in non-commutative 
space, replacing ordinary products by star products etc.

It is sometimes useful to express the self duality condition (\ref{consistency}) or
(\ref{variation}) in a different form that is more compact and may lead to generalisations. 
By introducing complex variables,
\be
M= F - iG
\label{complex}
\ee
this condition can be put in the form,
\be
\frac{\lambda}{4} M(\,^*M)^* = -\frac{\partial L}{\partial \theta}\delta\theta
\label{variation1}
\ee
where $M^*$ is the complex conjugate of $M$. For the Born-Infeld type theory, the right
side of the equation is zero. Schroedinger{\footnote{
For a review of Schroedinger's work, see \cite{GZ1}}}
 used these variables to discuss his formulation
of the Born-Infeld theory, the advantage being that it was {\it{manifestly}} 
covariant under the duality rotations $M\rightarrow M e^{i\lambda}$.

In our formulation it is possible to rephrase the self duality condition without the
need of introducing any complex variables. The idea is to redefine variables so that
these have a similar transformation property as the parameter $\theta$, given by 
(\ref{matrix1}). The new variables are,
\be
M_\pm = F \pm \,^*G
\label{pm}
\ee
which transform as,
\be
M_\pm\rightarrow \cos\lambda M_\pm \mp \sin\lambda\,^*M_\pm
\label{pmchange}
\ee
The self duality condition now takes the form.
\be
\frac{\lambda}{4}M_+\, ^*M_- = -\frac{\partial L}{\partial \theta}\delta\theta
\label{selfduality}
\ee
It may be worthwhle to pursue a Schroedinger-like analysis for non-commutative electrodynamics
with these variables. However, even in the present context they are useful, as will be 
illustrated later.

\bigskip

{\bf{Duality as a Legendre transformation}}

\bigskip

It is known \cite{GZ1, review} that duality transformations are Legendre transformations.
Also, any system that solves the self duality condition (\ref{consistency}) or 
(\ref{variation}) will be automatically invariant under the Legendre transformations.
Here we show that there are systems which violate the self duality condition but are
nevertheless invariant under the Legendre transformations. The situation is analogous
to what has already been discussed, namely, the self duality condition is sufficient,
but not necessary, to ensure duality under discrete transformations. Similarly, self
duality condition is sufficient, though not necessary, to interpret duality as a Legendre
transformation. Indeed, both the generalised systems (\ref{BItype}) and
(\ref{NCEMtype}) satisfy this property although in general they do not solve the self
duality condition. The explicit proof will be shown for (\ref{NCEMtype}), while the 
other just follows from identical steps.

The lagrangian $L(F, \theta)$
 (\ref{NCEMtype}) is expressed by its Legendre transformed version as,
\be
L(F, \theta, F_D)= L(F, \theta) -\frac{1}{2}F\,^*F_D\,\,\,;\,\,\, F_D^{\mu\nu}
=\partial^\mu A_D^\nu -\partial^\nu A_D^\mu
\label{l1}
\ee
where $F$ is now regarded as an unconstrained antisymmetric field, $A_D$ is a Lagrange
multiplier and $F_D$ is the dual electromagnetic field. This model is equivalent 
to the original one. To see this note that the equation of motion for $A_D$ imposes
$\partial_\mu\,^*F^{\mu\nu} =0$ so that the second term is a total derivative and the
two lagrangians get identitfied. The dual version is now obtained by 
eliminating $F$ in favour of $F_D$, using the equations
of motion for $F$, which yields,
\be
^*F_D = 2\frac{\partial L}{\partial F} = \,^*G
\label{l2}
\ee
where the last equality follows on using the basic definition (\ref{dual}). It is now 
possible to invert the relation (\ref{ncdual}) to obtain a solution for $F$
in terms of $G$, which, up to the order we are interested in, is given by,
\be
F= -\,^*G - 4(\theta.\,^*G)\,^*G - 
2\theta \,^*G^2 - 2b(\theta\,^*G\,^*G + \,^*G\theta\,^*G
+\,^*G\,^*G\theta)
\label{l3}
\ee
Putting this back into (\ref{l1}), one finds the dual lagrangian,
\be
L_D(G, \theta)= \frac{1}{4}\,^*G^2 + (\theta\,^*G)\,^*G^2 
- b(\theta\,^*G\,^*G\,^*G)^\mu\,\,_\mu
\label{l4}
\ee
Using $^*G^2 = -G^2$ and identities already mentioned, this simplifies to the desired form,
\be
L_D(G, \theta) = -\frac{1}{4} G^2 - (\,^*\theta G)G^2  +b (\,^*\theta GGG)^\mu\,\,_\mu
\label{l5}
\ee
This dual lagrangian precisely follows from (\ref{NCEMtype}) by using the discrete
duality transformations $(F\rightarrow G, \theta\rightarrow \,^*\theta)$. It shows
how duality transformations act as Legendre transformations. Note that the result 
is valid for any value of the arbitrary parameter $(b)$.

One may wonder, recalling that in the general proof \cite{GZ1, review}, self duality condition 
is used to show invariance under the Legendre transformation, how did this feature survive
even though the self duality condition is violated for a general value of the parameter.
For instance, in the Born-Infeld theory, self duality is used to prove that
the expression $(L-\frac{1}{4}F\,^*G)$ is invariant, after which the proof goes through.
In the present case it is even simpler because, as will soon be shown, the 
extra terms in (\ref{NCEMtype}) are already invariant so that the usual
considerations can be directly applied to the Maxwell piece and the proof
follows trivially.
To see this, 
observe that the product $M_+ M_-$ is invariant under the continuous transformations
(\ref{pmchange}). From (\ref{pm}) and (\ref{ncdual1}), it follows that,
\be
M_+ M_- = F^2 + G^2 = -12 (\theta F)(F^2) + 12 b  (\theta FFF){^\m\,\, _\m}
\label{product}
\ee
which is proportional to the correction terms in (\ref{NCEMtype}). Thus these terms
are invariant under the continuous duality rotations and the only change comes from the
standard Maxwell piece. 
This illuminates the reason for the duality
transformation acting as a Legendre transformation in the model defined by 
(\ref{NCEMtype}) and also concludes the discussion for this part.

The results of our analysis may be used for cases where quantum efects have been 
included. For instance, the one-loop effective action of ordinary QED is given by,
\be
{\cal L}_{eff} = 
 -\frac{1}{4} F^2 - \frac{\alpha^2}{36m^4}(F^2)^2 +\frac{7\alpha^2}{90m^4}
(FFFF){^\m\,\, _\m}
\label{qed}
\ee
where $\alpha$ is the fine structure constant and $m$ is the mass of the electron.
The last two terms are the Euler-Heisenberg corrections. This form has a close
resemblance to the ordinary Born-Infeld or the non-commutative Maxwell lagrangians.
This one-loop effect has interesting consequences. In particular,
 it was shown \cite{Adler} that there
was a modification in the dispersion relation in the presence of a constant external
magnetic field, leading to the electomagnetic birefringence phenomenon. In the present
context, since the ratio of the coefficients of the corrections terms is different
from four, there will be no continuous duality symmetry. Thus the one-loop effect 
destroys this property which is otherwise manifested in the pure Maxwell theory.
However, the symmetry under discrete duality transformations will still be retained,
since it is independent of the ratio.

As is known \cite{Shore},
just as the theory defined by (\ref{qed}) leads to modified dispersion
relations, the same is also true for the Born-Infeld theory. In both these cases,
there is subluminal propagation; i.e. the photons travel at a speed that is less
that the speed of light. A similar effect occurs for the non-commutative Maxwell
theory (\ref{NCEM}), although here superluminal propagation is possible\cite{J, C,
ABT}. The clash
with causality is avoided by realising that Lorentz covariance in (\ref{NCEM}) is
broken since $\theta^{\mu\nu}$, contrary to $F^{\mu\nu}$, does not transform like
a tensor. Thus the presence of $\theta^{\mu\nu}$ makes a significant difference in
the dispersion relations.
However as far as duality properties are concerned, the effects of 
 $\theta^{\mu\nu}$ and  $F^{\mu\nu}$ look very similar. Both the generalised Born-Infeld
type and generalised non-commutative Maxwell type theories 
(the latter following from the former by formally
replacing one of the $F$'s by a $\theta$)
revealed identical 
features under duality symmetry, discrete or continuous. In general, 
only the discrete duality symmetry was preserved. The continuous symmetry was
present provided the ratio of the coefficients in the nonlinear terms
was four (with the proper sign)
 and in that case the 
generalised lagrangians exactly reduced to the standard Born-Infeld or non-commutative
Maxwell theories. It may be mentioned that the occurrence of even discrete duality
symmetry is sufficient to extract new solutions from known solutions, as was recently
discussed \cite{ABT} for non-commutative Maxwell type theories (\ref{NCEMtype}).
 Interestingly, duality invariance under
Lagendre transformations was preserved  for the general case and not restricted to
any particular ratio, as dictated by the self duality condition. For future 
possibilities we mention the extension of this analysis to higher orders. Also,
our investigations reveal that a more detailed enquiry into the connection between
self duality condition and Legendre transformations is desirable.

\bigskip

{\Large{\bf{Acknowledgements}}}

\bigskip

This work was supported by a grant from the SungKyunKwan University. I also thank
the members of the theory group, physics department, for their kind hospitality.

\end{document}